\documentclass{article}

\usepackage{spconf,amsmath,graphicx,hyperref}
\usepackage{cite}
\usepackage{amsmath,amssymb,amsfonts}
\usepackage{algorithmic}
\usepackage{graphicx,color}
\usepackage{textcomp}
\usepackage{xcolor}
\usepackage{enumerate}
\usepackage[shortlabels]{enumitem}
\usepackage{hyperref}
\hypersetup{hidelinks=true}
\usepackage{algorithm,algorithmic}
\def\BibTeX{{\rm B\kern-.05em{\sc i\kern-.025em b}\kern-.08em
		T\kern-.1667em\lower.7ex\hbox{E}\kern-.125emX}}
\AtBeginDocument{\definecolor{tmlcncolor}{cmyk}{0.93,0.59,0.15,0.02}\definecolor{NavyBlue}{RGB}{0,86,125}}

\usepackage{soul, xcolor}

\usepackage{booktabs}
\usepackage{cite}
\usepackage{graphicx}
\usepackage{psfrag}
\usepackage{url}
\usepackage{amsmath}
\usepackage{array}
\usepackage{amssymb}
\usepackage{mathtools}
\usepackage{amsfonts}
\usepackage{graphicx}
\usepackage{epstopdf}
\usepackage{algorithm}
\usepackage{cite,algorithm,algorithmic,amsmath,amssymb,amsthm,empheq,mhsetup}
\usepackage{algorithmic}
\newtheorem{proposition}{Proposition}

\newtheorem{remark}{Remark}

\usepackage{setspace}
\usepackage{amscd}
\usepackage{mathrsfs}
\usepackage{epsfig}
\usepackage{color}
\usepackage{textcomp}
\usepackage{float}
\usepackage{pgf}
\usepackage{pgfplots}
\usepackage{tikz}
\usetikzlibrary{spy}
\usepackage{eqparbox}
\usepgfplotslibrary{groupplots}
\usetikzlibrary{decorations.pathreplacing,calligraphy}
\usetikzlibrary{shapes.geometric}

\usepackage{lipsum}  

\usepackage{glossaries}
\usepackage{glossaries-extra}

\usepackage{gensymb}

\usepackage{soul}
\usepackage{flushend}

\setabbreviationstyle[acronym]{long-short}
\newacronym{fwa}{FWA}{fixed wireless access}
\newacronym{mimo}{MIMO}{multiple input multiple output}
\newacronym{5g}{5G}{fifth generation}
\newacronym{iid}{i.i.d}{independent and identically distributed}
\newacronym{bs}{BS}{base station}
\newacronym{ue}{UE}{user equipment}
\newacronym{ap}{AP}{access point}
\newacronym{cpu}{CPU}{central processing unit}
\newacronym{upa}{UPA}{uniform planar array}
\newacronym{los}{LoS}{line-of-sight}
\newacronym{awgn}{AWGN}{additive white Gaussian noise}
\newacronym{isp}{ISP}{internet service provider}
\newacronym{mmse}{MMSE}{minimum mean square error}
\newacronym{mse}{MSE}{mean square error}
\newacronym{zf}{ZF}{zero-forcing}
\newacronym{mrc}{MRC}{maximum ratio combiner}
\newacronym{mrt}{MRT}{maximum ratio transmission}
\newacronym{se}{SE}{spectral efficiency}
\newacronym{csi}{CSI}{channel state information}
\newacronym{snr}{SNR}{signal-to-noise ratio}
\newacronym{sinr}{SINR}{signal-to-interference-plus-noise ratio}
\newacronym{tdd}{TDD}{time-division duplexing}
\newacronym{ul}{UL}{uplink}
\newacronym{dl}{DL}{downlink}
\newacronym{cli}{CLI}{cross-link interference}
\newacronym{inai}{InAI}{inter-AP interference}
\newacronym{inui}{InUI}{inter-user interference}

\DeclareMathOperator{\EEE}{\mathbb{E}}
\DeclareMathOperator{\var}{\mathrm{Var}}

\DeclareMathOperator{\F}{\mathbf{F}}

\DeclareMathOperator{\vv}{\mathbf{v}}
\DeclareMathOperator{\w}{\mathbf{w}}

\DeclareMathOperator{\h}{\mathbf{h}}

\DeclareMathOperator{\R}{\mathcal{R}}

\DeclareMathOperator{\CN}{\mathcal{CN}}

\DeclareMathOperator{\II}{\mathbf{I}}

\DeclareMathOperator{\x}{\mathbf{x}}

\DeclareMathOperator{\y}{\mathbf{y}}

\DeclareMathOperator{\g}{\mathbf{g}}

\DeclareMathOperator{\0}{\mathbf{0}}

\makeatletter
\newcommand\fs@spaceruled{\def\@fs@cfont{\bfseries}\let\@fs@capt\floatc@ruled
	\def\@fs@pre{\vspace{0.5\baselineskip}\hrule height.7pt depth0pt \kern2pt}%
	\def\@fs@post{\kern2pt\hrule\relax}%
	\def\@fs@mid{\kern2pt\hrule\kern2pt}%
	\let\@fs@iftopcapt\iftrue}

\newsavebox\myboxA
\newsavebox\myboxB
\newlength\mylenA
\newcommand*\mybar[2][0.66]{
	\sbox{\myboxA}{$\m@th#2$}
	\setbox\myboxB\null
	\ht\myboxB=\ht\myboxA
	\dp\myboxB=\dp\myboxA
	\wd\myboxB=#1\wd\myboxA
	\sbox\myboxB{$\m@th\overline{\copy\myboxB}$}
	\setlength\mylenA{\the\wd\myboxA}
	\addtolength\mylenA{-\the\wd\myboxB}
	\ifdim\wd\myboxB<\wd\myboxA
	\rlap{\hskip 0.5\mylenA\usebox\myboxB}{\usebox\myboxA}
	\else
	\hskip -0.5\mylenA\rlap{\usebox\myboxA}{\hskip 0.5\mylenA\usebox\myboxB}
	\fi}

\makeatother

\usepackage{tkz-euclide}
\tikzset{
	bs/.pic = {                                      
		\draw[line width = 1pt,-round cap] (0,0.4\R) -- (-0.2\R,-0.2\R);
		\draw[line width = 1pt,-round cap] (0,0.4\R) -- (0.2\R,-0.2\R);
		\draw[line width = 1pt] (-0.2\R,-0.2\R) -- (0.133\R,0);
		\draw[line width = 1pt] (0.2\R,-0.2\R) -- (-0.133\R,0);
		\draw[line width = 1pt,-round cap] (-0.133\R,0) -- (0.067\R,0.2\R);
		\draw[line width = 1pt,-round cap] (0.133\R,0) -- (-0.067\R,0.2\R);
		\draw[line width = 1pt,-round cap] (-0.213\R,0.4\R) -- (0.213\R,0.4\R);
		\draw[line width = 1pt,-round cap] \foreach \x in {-0.21, -0.14,...,0.22} {(\x\R,0.4\R) -- (\x\R,0.47\R)};
		\node (dim) at (0,0)  [align=center,minimum width=0.5\R,minimum height=1\R] {};
	},
}

\tikzset{
	user/.pic = {     
		\draw[rounded corners=0.02\R] (-0.09\R,-0.17\R) rectangle (0.09\R,0.17\R) ;                     
		\draw[fill=gray] (-0.08\R,-0.12\R) rectangle (0.08\R,0.12\R);                          
		\draw[rounded corners=0.005\R] (-0.04\R,0.14\R) rectangle (0.04\R,0.15\R);                     
		\draw (0,-0.14\R) circle (0.012\R) ; 
		\node (dim) at (0,0)  [align=center,minimum width=0.2\R,minimum height=0.3\R] {};
	}
}

\tikzset{
	repeater/.pic = {     
		\draw[rounded corners=0.02\R] (-0.25\R,-0.15\R) rectangle (0.25\R,0.15\R) ;                     
		\draw[fill=white] (-0.24\R,-0.14\R) rectangle (0.24\R,0.14\R);
		\draw[line width = 1pt] (0.24\R,0\R) -- (0.45\R,0\R);
		\draw[line width = 1pt] (-0.24\R,0\R) -- (-0.45\R,0\R);     
		\draw (-0.47\R,0\R) circle (0.03\R) ; 
		\draw (0.47\R,0\R) circle (0.03\R) ;
		\node (dim) at (0,0)  [align=center,minimum width=0.2\R,minimum height=0.3\R] {};
	}
}

\title{Is Repeater-Assisted Massive MIMO Compatible with Dynamic TDD?}
\name{Martin Andersson, Anubhab Chowdhury, and Erik G. Larsson \thanks{This work was partially supported by the Wallenberg AI, Autonomous Systems and Software Program (WASP) funded by the Knut and Alice Wallenberg Foundation, and partially supported by ELLIIT.}}
\address{Department of Electrical Engineering (ISY), Link\"{o}ping University, 581 83 Link\"{o}ping, Sweden \\ Email: \texttt{\{martin.b.andersson, anuch87, erik.g.larsson\}@liu.se}}

\allowdisplaybreaks
\begin{document}
\ninept
\maketitle
\begin{abstract}
    We present a framework for joint amplification and phase shift optimization of the repeater gain in dynamic time-division duplex (TDD) repeater-assisted massive MIMO networks. Repeaters, being active scatterers with amplification and phase shift, enhance the received signal strengths for users. However, they inevitably also amplify undesired noise and interference signals, which become particularly prominent in dynamic TDD systems due to the concurrent downlink (DL) and uplink (UL) transmissions, introducing cross-link interference among access points and users operating in opposite transmit directions. 
    This causes a non-trivial trade-off between amplification of desired and undesired signals.
    To underpin the conditions under which such a trade-off can improve performance, we first derive DL and UL spectral efficiencies (SEs), and then develop a repeater gain optimization algorithm for SE maximization. 
    Numerically, we show that our proposed algorithm successfully calibrates the repeater gain to amplify the desired signal while limiting the interference.
\end{abstract}

\vspace{-\baselineskip}

\section{INTRODUCTION}

\vspace{-2mm}

\label{sec:intro}

Repeaters are devices that instantaneously receive, amplify, phase shift, and retransmit signals, effectively acting as
\emph{active scatterers} in the propagation environment.
They can be deployed on a large scale to enhance coverage
for multiple-input multiple-output~(MIMO) wireless
systems~\cite{Willhammar2025CM, Garcia_TWC, Larsson2024WCL,Tsai_TWC, Sergiy_WCL, Carvalho2025CSM, Alp2025arXiv, Bai2026TWC, Yiming_PIMRC}. In~\cite{Willhammar2025CM}, it was demonstrated that repeater-assisted single-cell massive MIMO can procure spectral efficiencies~(SEs) similar to distributed MIMO, while obviating huge front-haul signaling overhead and the necessity of precise phase alignment among access points~(APs). Furthermore, the authors in~\cite{RIS_Makki_IEEE_Acess} demonstrated that to procure similar performance as with repeaters, reflecting intelligent surface~(RIS)-assisted systems require a significantly larger number of reflecting elements, making its deployment more challenging compared to repeaters.
The easy-to-implement features of these repeaters have led to
their inclusion in the 5G NR standards since 3GPP Release
$18$~\cite{3gpp}.

We note that the papers cited above consider \emph{static} time-division
duplexing~(TDD), which is not always spectrally efficient in scenarios with
heterogeneous \gls{dl} and \gls{ul} traffic
demands~\cite{Andersson2023Asilomar,El2011GLOBECOM}. On the contrary, \emph{dynamic TDD} serves \gls{ul} and
\gls{dl} users concurrently by allowing spatially separated
half-duplex APs to be scheduled in either DL or UL on the same time-frequency
resources, offering improved spectral and energy efficiency compared
to traditional static TDD-based MIMO systems~\cite{Chowdhury_WCL, Zhu2021CL, Mohammadi2023JSAC, KimCST, Jiamin_TWC}.

However, due to the concurrent \gls{dl} and \gls{ul} communication in dynamic TDD
systems, a repeater inevitably amplifies and forwards the cross-link
interference~(CLI), i.e., the \gls{dl} AP to \gls{ul} AP~(inter-AP)
interference and \gls{ul} user to \gls{dl} user~(inter-user)
interference, in addition to the desired \gls{dl} and \gls{ul} data
signals. Thus, it is critical to analyze the conditions
under which repeaters are compatible with dynamic TDD, i.e., when
substantial benefits can be procured by deploying repeaters in dynamic
TDD systems, while ensuring that the CLI amplification does not
overwhelm the desired signal gains.

\textbf{Contributions:} In this paper, we provide a
framework for repeater gain optimization in dynamic TDD systems. In contrast to existing works on repeater gain optimization (in static TDD) that only consider amplification~\cite{Alp2025arXiv,Bai2026TWC}, we account for a complex-valued repeater gain which also includes a phase shift. We provide the following specific results: 1) We derive closed-form
achievable DL and UL SEs for a repeater-assisted massive MIMO
network operating with \emph{dynamic} TDD, taking the noise and all sources of
interference into consideration; that is, multi-user, inter-AP,
inter-user, and the corresponding amplified interference terms caused
by the repeater. 2) We propose a method to jointly optimize the
amplification and phase shift of the \emph{complex-valued} repeater gain for
SE maximization, using an iterative framework that alternates between
optimizing the repeater gain and updating the precoding and combining
vectors. 3) Numerical results prove that our proposed repeater gain
optimization algorithm effectively improves the SE in the direction (DL/UL) for which it is optimized, but can slightly reduce the SE in the opposite direction in certain cases; however, the gains are much more significant than the losses. Further, we observe that substantial SE improvements can be obtained by optimizing the repeater phase shift. 

\vspace{-\baselineskip}

\section{SYSTEM MODEL}
\label{sec:syst}

\vspace{-2mm}

\begin{figure}[t]
    \centering
    \begin{tikzpicture}
		\newdimen\R
		\R=1.25cm
		
		\node[
		regular polygon,
		regular polygon sides=6,
		draw,
		thick,
		rotate=90,
		inner sep=1.25cm,
		dashed,
		] (c1) at (0,0) {};
		
		\node[
		regular polygon,
		regular polygon sides=6,
		draw,
		thick,
		rotate=90,
		inner sep=1.25cm,
		dashed,
		] (c2) at (3.535,0) {};
		
		\draw (0,0) pic (dlap) {bs} node [xshift=-8mm,yshift=5mm] {\scriptsize DL AP} ;		
		\draw (3.535,0) pic (ulap) {bs} node [xshift=8mm,yshift=5mm] {\scriptsize UL AP} ;
		
		\draw (1.7675,0) pic (rep) {repeater} node [yshift=5mm,fill=white,inner sep=1.25pt] {\scriptsize Repeater} ;
		
		\draw (0,-1.25) pic (dlue) {user} node [xshift=-9mm,yshift=-2mm,fill=white,inner sep=2pt] {\scriptsize DL user $k$} ;
		\draw (3.55,-1.25) pic (ulue) {user} node [xshift=9mm,yshift=-2mm,fill=white,inner sep=2pt] {\scriptsize UL user $j$} ;

		\draw[black,thick,->] ([yshift=4mm]dlapdim.south) to node [align=center,midway,xshift=-4mm,yshift=0mm]{\small $\g_{\text{d},k}^T$} ([yshift=1mm]dluedim.north);
		
		\draw[black,thick,<-] ([yshift=4mm]ulapdim.south) to node [align=center,midway,xshift=4mm,yshift=0mm]{\small $\g_{\text{u},j}$} ([yshift=1mm]uluedim.north);
	
		\draw[black,thick,<-] ([xshift=1mm]dluedim.east) to node [align=center,midway,xshift=1mm,yshift=-3mm]{\small $h_{\text{d},k}$} ([xshift=4mm,yshift=-1mm]repdim.east);
		
		\draw[black,thick,<-]([xshift=-4mm,yshift=-1mm]repdim.west) to node [align=center,midway,xshift=1mm,yshift=-3mm]{\small $h_{\text{u},j}$} ([xshift=-1mm]uluedim.west);
		
		\draw[black,thick,->] ([xshift=5.75mm]repdim.east) to node [align=center,midway,xshift=0mm,yshift=2mm]{\small $\h_{\text{u}}$} (ulapdim.west);
		
		\draw[black,thick,<-] ([xshift=-5.75mm]repdim.west) to node [align=center,midway,xshift=0mm,yshift=2.5mm]{\small $\h_{\text{d}}^T$} (dlapdim.east);
		
		\draw[red,thick,->] (dlapdim.north) to[out=45,in=135] node [align=center,midway,xshift=0mm,yshift=2mm]{\small $\F$} (ulapdim.north) ;
		
		\draw[red,thick,<-] ([yshift=-1mm]dluedim.south) to[out=-45,in=-135] node [align=center,midway,xshift=0mm,yshift=2mm]{\small $f_{kj}$} ([yshift=-1mm]uluedim.south) ;

        \draw[blue,thick,-] ([xshift=-3.2mm,yshift=-0.125mm]repdim.north) to node [align=center,midway,fill=white,inner sep=0.5pt,xshift=0mm,yshift=-1.25mm]{\small $\alpha_\text{r}$} ([xshift=3.2mm,yshift=-0.125mm]repdim.north) ;

        \draw[blue,thick,->]([xshift=3.05mm]repdim.north) to node {}([xshift=3.05mm,yshift=-2mm]repdim.north);
        
        \draw[blue,thick,-]([xshift=-3.05mm]repdim.north) to node {}([xshift=-3.05mm,yshift=-2mm]repdim.north);
        
    \end{tikzpicture}
    \vspace{-5mm}
    \caption{ The two-cell repeater-assisted massive MIMO system operating with dynamic TDD. The CLI channels are denoted by red arrows. We let $\alpha_\text{r}$ be the repeater gain of the forward path.
    }
    \label{Fig:system}
    \vspace{-5mm}
\end{figure}
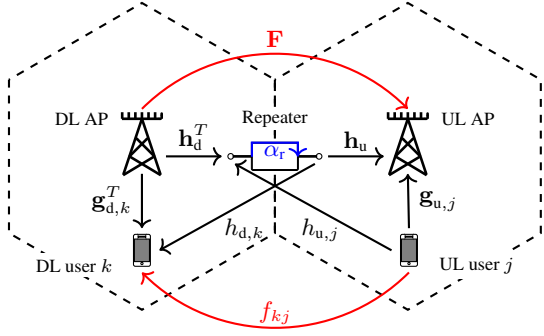

We consider a repeater-assisted massive MIMO network with dynamic TDD
operation. Specifically, we focus on the case with two cells, each
containing one $M$-antenna AP. One of the two cells operates in DL, and
the other cell operates in UL, on the same time-frequency resources.
Hereafter, we refer to the AP in the cell operating in DL and UL as
the DL and UL AP, respectively. The UL AP communicates with $J$
single-antenna UL users within its cell, whereas the DL AP serves $K$
single-antenna DL users. The APs are located in the cell centers, and
the DL and UL users are arbitrarily distributed within their
respective cells. Further, there is one dual-antenna repeater present
in the network; it can be located in any of the two cells. A sketch of
the system model is provided in Fig. \ref{Fig:system}.

\begin{remark}
    We note that a dual-antenna repeater has a forward and a reverse
    path, and the corresponding gains need to be pre-calibrated for reciprocity so
    that the repeater appears transparent to the wireless propagation
    environment in TDD operation~\cite{Larsson2024WCL}. The repeater's
    forward and reverse paths are scheduled as per the TDD pattern of
    the APs. However, dynamic TDD systems also include a phase with concurrent DL and UL transmissions~(where CLI appears), apart from the usual dedicated DL and UL phases~(as in static TDD). In the concurrent DL-UL phase, which is of concern in this work, the repeater can choose either of the paths. Our subsequent analysis holds irrespective of whether the repeater is receiving via the
    forward or the reverse path; the repeater will amplify and forward the CLI in any case. Thus, without loss of generality, we can assume that $\alpha_\text{r}\in\mathbb{C}$ is the forward path gain.
\end{remark}
In accordance with Fig. \ref{Fig:system}, we define the following channels. First, $\{ \g_{\text{d},k} \}$ are the channels from each DL user $k$ to the DL AP. Similarly, $\{ \g_{\text{u},j} \}$ are the channels from each UL user $j$ to the UL AP. Further, $\h_{\text{d}}$ and $\h_{\text{u}}$ are the channels from the repeater to the DL AP and the UL AP, respectively. The channels from the repeater to the DL users are $\{ h_{\text{d},k} \}$ and to the UL users are $\{ h_{\text{u},j} \}$. Finally, the inter-AP channel is denoted $\F$, and the inter-user channels between each DL user $k$ and each UL user $j$ are denoted by $\{ f_{kj} \}$. We assume that all channels are perfectly known and that this knowledge is available everywhere in the network.

\vspace{-\baselineskip}

\subsection{\gls{dl} and \gls{ul} Data Transmission}

\vspace{-1mm}

We now define the data transmission models in DL and UL, latter used to calculate achievable DL and UL SEs. Let $\{ s_{\text{d},k}\}$ denote the data symbols intended for the $K$ DL users and $\{ s_{\text{u},j} \}$ be the data symbols transmitted by the $J$ UL users. All data symbols have zero mean, unit variance, and are mutually uncorrelated. Further, let $\{ \eta_{\text{d},k} \}$ and $\{ \eta_{\text{u},j} \}$ be the power control coefficients for the DL and UL users, respectively, where $0 \leq \eta_{\text{d},k}, \ \forall k$ and $0 \leq \eta_{\text{u},j} \leq 1, \ \forall j$. Then, the signal transmitted by the DL AP can be expressed as $\x_{\text{d}} = \sum_{k=1}^K \sqrt{\eta_{\text{d},k}} \w_k s_{\text{d},k}$, where $\{ \w_k \}$ are the precoding vectors applied by the DL AP for each DL user. In addition to being non-negative, the DL power control coefficients must be chosen such that the DL power constraint $\EEE \{ \x_{\text{d}}^H \x_{\text{d}} \} \leq 1$ is fulfilled, which can be equivalently formulated as $\sum_{k=1}^K \eta_{\text{d},k} \w_k^H \w_k \leq 1.$
Further, each UL user $j$ transmits the signal $x_{\text{u},j} = \sqrt{\eta_{\text{u},j}} s_{\text{u},j}$.

Recall that, in dynamic TDD, both the DL AP and the UL users transmit their data simultaneously. Thus, the signal impinging on the repeater is given by
\begin{align}
    \label{eq:y_in}
    \nonumber
    y_\text{r}^{\text{in}} & = \sqrt{\rho_\text{d}} \h_{\text{d}}^T \x_{\text{d}} + \sum\nolimits_{j=1}^J \sqrt{\rho_\text{u}} h_{\text{u},j} x_{\text{u},j}
    \\
    & = \sum\nolimits_{k=1}^K \sqrt{\rho_\text{d} \eta_{\text{d},k}} \h_{\text{d}}^T \w_k s_{\text{d},k} + \sum\nolimits_{j=1}^J \sqrt{\rho_\text{u} \eta_{\text{u},j}} h_{\text{u},j} s_{\text{u},j}.
\end{align}
Here, $\rho_\text{d}$ and $\rho_\text{u}$ denote the DL and UL signal-to-noise ratios (SNRs), respectively.
The input is scaled by the complex-valued repeater gain $\alpha_\text{r} = r_\text{r} e^{i \phi_\text{r}}$, where $r_\text{r} = |\alpha_\text{r}|$ is the amplification gain and $\phi_\text{r} = \angle \alpha_\text{r}$ is the phase shift. Then, the repeater radiates the signal
\begin{align}
	\label{eq:yrout}
	y_\text{r}^{\text{out}} = \alpha_\text{r} ( y_\text{r}^{\text{in}} + w_\text{r}),
\end{align}
where $w_\text{r} \sim \CN(0, \sigma_\text{r}^2)$ is the additive noise at the repeater. The amplification gain must be chosen such that \eqref{eq:yrout} does not exceed the output power constraint $\EEE \{ |y_\text{r}^{\text{out}}|^2 \} = r_\text{r}^2 (P_\text{in} + \sigma_\text{r}^2) \leq P_\text{max}$ \cite{Alp2025arXiv}, i.e.,
\begin{align}
    \label{eq:P_max}
    0 \leq r_\text{r} \leq \sqrt{P_\text{max}/(P_\text{in} + \sigma_\text{r}^2)},
\end{align}
where we defined the power of the repeater input as $P_\text{in} = \EEE \{|y_\text{r}^{\text{in}}|^2\} = \sum_{k=1}^K \rho_\text{d} \eta_{\text{d},k} |\h_{\text{d}}^T \w_k|^2 + \sum_{j=1}^J \rho_\text{u} \eta_{\text{u},j} |h_{\text{u},j}|^2 $. We note that in scenarios with multiple interacting repeaters, the amplification gain is also limited by stability constraints \cite{Bai2026TWC}.

\vspace{-\baselineskip}

\subsection{\gls{dl} and \gls{ul} Spectral Efficiencies}

\vspace{-1mm}

First, we calculate an achievable SE for each DL user. To this end, write the signal received at the $k$th DL user as
\begin{align}
    \label{eq:y_d}
    \nonumber
    y_{\text{d},k} & = \sqrt{\rho_\text{d}} \g_{\text{d},k}^T \x_{\text{d}} + h_{\text{d},k} y_\text{r}^{\text{out}} + \sum\nolimits_{j=1}^J \sqrt{\rho_\text{u}} f_{kj} x_{\text{u},j} + w_{\text{d},k}
    \\
    \nonumber
    & = \sum\nolimits_{k'=1}^K \sqrt{ \rho_\text{d} \eta_{\text{d},k'}} \g_{\text{d},k}^T \w_{k'} s_{\text{d},k'} + \alpha_\text{r} h_{\text{d},k} ( y_\text{r}^{\text{in}} + w_\text{r})
    \\
    \nonumber
    & \quad + \sum\nolimits_{j=1}^J \sqrt{\rho_\text{u} \eta_{\text{u},j}} f_{kj} s_{\text{u},j} + w_{\text{d},k}
    \\
    \nonumber
    & = \sqrt{ \rho_\text{d} \eta_{\text{d},k}} \mybar\g_{\text{d},k}^T \w_k s_{\text{d},k} + \sum\nolimits_{k'=1,k' \neq k}^K \sqrt{ \rho_\text{d} \eta_{\text{d},k'}} \mybar\g_{\text{d},k}^T \w_{k'} s_{\text{d},k'}
    \\
    & \quad + \sum\nolimits_{j=1}^J \sqrt{\rho_\text{u} \eta_{\text{u},j}} \mybar{f}_{kj} s_{\text{u},j} + \mybar{w}_k,
\end{align}
where $w_{\text{d},k} \sim \CN(0, 1)$ is the additive receiver noise at DL user $k$. In \eqref{eq:y_d}, for each DL user $k$, we also introduced the composite DL channel (direct link and via the repeater) $\mybar\g_{\text{d},k} = \g_{\text{d},k} + \alpha_\text{r} h_{\text{d},k} \h_{\text{d}}$, and the corresponding interference and noise terms $\mybar{f}_{kj} = f_{kj} + \alpha_\text{r} h_{\text{d},k} h_{\text{u},j}$ and $\mybar{w}_k = \alpha_\text{r} h_{\text{d},k} w_\text{r} + w_{\text{d},k}$. From \eqref{eq:y_d}, we derive an achievable SE for each DL user.
\begin{proposition}
    \label{thm:SE_DL}
    An achievable SE for each DL user $k$ is given by $\mathtt{SE}_{\text{d},k} = \log_2 \left( 1 + \mathtt{SINR}_{\text{d},k} \right)$,
    where
    \begin{align}
        \label{eq:SINR_d}  
        \nonumber
        \mathtt{SINR}_{\text{d},k} & = \rho_\text{d} \eta_{\text{d},k} |\mybar\g_{\text{d},k}^T \w_{k}|^2 \Bigg/
        \Bigg( \sum_{k'=1,k' \neq k}^K \rho_\text{d} \eta_{\text{d},k'} |\mybar\g_{\text{d},k}^T \w_{k'}|^2
        \\
        & \qquad  + \sum\nolimits_{j=1}^J \rho_\text{u} \eta_{\text{u},j} |\mybar{f}_{kj}|^2+ \sigma_\text{r}^2 |\alpha_\text{r} h_{\text{d},k}|^2 + 1 \Bigg).
    \end{align}
\end{proposition}
\begin{proof}
    Since all channels are perfectly known and available everywhere, the received signal at user $k$ in \eqref{eq:y_d} is practically an observation of the intended data symbol $s_{\text{d},k}$ over the deterministic channel $\sqrt{ \rho_\text{d} \eta_{\text{d},k}} \mybar\g_{\text{d},k}^T \w_{k}$. The remaining terms in \eqref{eq:y_d}, i.e. $e_{\text{d},k} = y_{\text{d},k} - \sqrt{ \rho_\text{d} \eta_{\text{d},k}} \mybar\g_{\text{d},k}^T \w_{k} s_{\text{d},k}$, constitute the effective non-Gaussian noise. Hence, using \cite[Corol. 1.3]{emil17}, the capacity is lower bounded by $\log_2 (1 + |\sqrt{ \rho_\text{d} \eta_{\text{d},k}} \mybar\g_{\text{d},k}^T \w_{k}|^2 / \var\{ e_{\text{d},k} \})$. Standard calculations give \eqref{eq:SINR_d}; we omit a detailed derivation due to space limitations.
\end{proof}

Next, we derive an achievable SE for each UL user. The signal received at the UL AP can be expressed as
\begin{align}
    \y_{\text{u}} & = \sum\nolimits_{j=1}^J \sqrt{\rho_{\text{u}}} \g_{\text{u},j} x_{\text{u},j} + \h_{\text{u}} y_\text{r}^{\text{out}} + \sqrt{\rho_\text{d}} \F \x_{\text{d}} + \w_\text{u},
\end{align}
where $\w_\text{u} \sim \CN(\0, \II_M)$ is the receiver noise at the UL AP. Let $\vv_j$ be the combining vector for the data stream of the $j$th UL user, yielding the soft estimate of $s_{\text{u},j}$ as $\vv_j^H \y_{\text{u}}$, which expands to
\begin{align}
    \label{eq:y_u}
    \nonumber
    \vv_j^H \y_{\text{u}} & = \sqrt{\rho_{\text{u}} \eta_{\text{u},j}} \vv_j^H \mybar\g_{\text{u},j} s_{\text{u},j} + \sum_{j'=1,j' \neq j}^J \sqrt{\rho_{\text{u}} \eta_{\text{u},j'}} \vv_j^H \mybar\g_{\text{u},j'} s_{\text{u},j'}
    \\
    & \quad + \sum\nolimits_{k=1}^K \sqrt{\rho_\text{d} \eta_{\text{d},k}} \vv_j^H \mybar\F \w_k s_{\text{d},k} + \vv_j^H \mybar\w_{\text{u}},
\end{align}
where for each UL user $j$ we introduced the composite UL channel $\mybar\g_{\text{u},j} = \g_{\text{u},j} + \alpha_\text{r} h_{\text{u},j} \h_{\text{u}}$, as well as the corresponding interference and noise terms $\mybar\F = \F + \alpha_\text{r} \h_{\text{u}} \h_{\text{d}}^T$ and $\mybar\w_{\text{u}} = \alpha_\text{r} w_\text{r} \h_{\text{u}} + \w_\text{u}$. From \eqref{eq:y_u}, we derive an achievable SE for each UL user.
\begin{proposition}
    \label{thm:SE_UL}
    An achievable SE for each UL user $j$ is given by $\mathtt{SE}_{\text{u},j} = \log_2 \left( 1 + \mathtt{SINR}_{\text{u},j} \right)$,
    where
    \begin{align}
        \label{eq:SINR_u}
        \nonumber
        & \mathtt{SINR}_{\text{u},j} = \rho_{\text{u}} \eta_{\text{u},j} |\vv_j^H \mybar\g_{\text{u},j}|^2 \Bigg/ \Bigg( \sum_{j'=1,j' \neq j}^J \rho_{\text{u}} \eta_{\text{u},j'} |\vv_j^H \mybar\g_{\text{u},j'}|^2 
        \\
        & \quad + \sum\limits_{k=1}^K \rho_\text{d} \eta_{\text{d},k} |\vv_j^H \mybar\F \w_k|^2 + \sigma_\text{r}^2 | \alpha_\text{r} \vv_j^H \h_{\text{u}}|^2 + \vv_j^H \vv_j \Bigg).
    \end{align}
\end{proposition}
\begin{proof}
    Similar to the proof of Proposition \ref{thm:SE_DL}, and hence omitted.
\end{proof}

\vspace{-\baselineskip}

\section{OPTIMIZING THE REPEATER GAIN}

\vspace{-2mm}

In this section, we develop an algorithm to find the repeater gain which maximizes the SE of one single DL or UL user. We do not account for the performance of the remaining users; developing such algorithms maximizing, e.g., the sum SE, is left for future work. We do this restriction as it allows for a rather simple optimization framework which lets us extract initial insights into the compatibility of dynamic TDD and repeater-assisted massive MIMO.

Maximizing the SE for a given user is equivalent to maximizing the corresponding SINR, under the power constraint \eqref{eq:P_max}. Note that both SINRs \eqref{eq:SINR_d} and \eqref{eq:SINR_u} are rational functions in the amplification gain $r_\text{r}$; specifically ratios of second-order polynomials on the form:
\begin{align}
    \label{eq:SINR_du}
    \mathtt{SINR}_{\text{d}/\text{u}, k/j} = \dfrac{a_{0,k/j} + a_{1,k/j} r_\text{r} + a_{2,k/j} r_\text{r}^2}{b_{0,k/j} + b_{1,k/j} r_\text{r} + b_{2,k/j} r_\text{r}^2},
\end{align}
where in the DL SINR \eqref{eq:SINR_d} we have
\begin{align}
    \label{eq:a0_DL}
    a_{0,k} & = \rho_\text{d} \eta_{\text{d},k} |\g_{\text{d},k}^T \w_k|^2,
    \\
    a_{1,k} & = 2 \rho_\text{d} \eta_{\text{d},k} \mathcal{R} \{ e^{-i \phi_\text{r}} h_{d,k}^* \g_{\text{d},k}^T \w_k \w_k^H \h_{\text{d}}^* \},
    \\
    a_{2,k} & = \rho_\text{d} \eta_{\text{d},k} |h_{\text{d},k} \h_{\text{d}}^T \w_k|^2,
    \\ 
    \label{eq:b0_DL}
    b_{0,k} & =\hspace*{-1em} \sum_{k'=1,k' \neq k}^K \rho_\text{d} \eta_{\text{d},k'} |\g_{\text{d},k}^T \w_{k'}|^2 + \sum_{j=1}^J \rho_\text{u} \eta_{\text{u},j} |f_{kj}|^2 + 1,
    \\
    \nonumber
    b_{1,k} & = \sum_{k'=1,k' \neq k}^K 2 \rho_\text{d} \eta_{\text{d},k'} \mathcal{R} \{ e^{-i \phi_\text{r}} h_{d,k}^* \g_{\text{d},k}^T \w_{k'} \w_{k'}^H \h_{\text{d}}^* \}
    \\
    & \qquad + \sum\limits_{j=1}^J 2 \rho_\text{u} \eta_{\text{u},j} \mathcal{R} \{ e^{-i \phi_\text{r}} f_{kj} h_{\text{d},k}^* h_{\text{u},j}^* \},
    \\
    \nonumber
    b_{2,k} & = \sum\nolimits_{k'=1,k' \neq k}^K \rho_\text{d} \eta_{\text{d},k'} |h_{\text{d},k} \h_{\text{d}}^T \w_{k'}|^2
    \\
    & \qquad + \sum\nolimits_{j=1}^J \rho_\text{u} \eta_{\text{u},j} |h_{\text{d},k} h_{\text{u},j}|^2 + \sigma_\text{r}^2 |h_{\text{d},k}|^2,
\end{align}
and in the UL SINR \eqref{eq:SINR_u} we have
\begin{align}
    \label{eq:a0_UL}
    a_{0,j} & = \rho_\text{u} \eta_{\text{u},j} |\vv_j^H \g_{\text{u},j}|^2,
    \\
    a_{1,j} & = 2 \rho_\text{u} \eta_{\text{u},j} \mathcal{R} \{ e^{-i \phi_\text{r}} h_{\text{u},j}^* \vv_j^H \g_{\text{u},j} \h_{\text{u}}^H \vv_j \},
    \\
    a_{2,j} & = \rho_\text{u} \eta_{\text{u},j} |h_{\text{u},j} \vv_j^H \h_\text{u}|^2,
    \\
    \nonumber
    b_{0,j} & = \sum\nolimits_{j'=1,j' \neq j}^J \rho_{\text{u}} \eta_{\text{u},j'} |\vv_j^H \g_{\text{u},j'}|^2 
    \\
    \label{eq:b0_UL}
    & \qquad + \sum\nolimits_{k=1}^K \rho_\text{d} \eta_{\text{d},k} |\vv_j^H \F \w_k|^2 + \vv_j^H \vv_j,
    \\
    \nonumber
    b_{1,j} & = \sum_{j'=1,j' \neq j}^J 2 \rho_{\text{u}} \eta_{\text{u},j'} \mathcal{R} \{ e^{-i \phi_\text{r}} h_{\text{u},j'}^* \vv_j^H \g_{\text{u},j'} \h_{\text{u}}^H \vv_j \}
    \\
    &  + \sum\nolimits_{k=1}^K 2 \rho_\text{d} \eta_{\text{d},k} \mathcal{R} \{ e^{-i \phi_\text{r}} \vv_j^H \F \w_k \w_k^H \h_\text{d}^* \h_\text{u}^H \vv_j \},
    \\
    \nonumber
    b_{2,j} & = \sum\nolimits_{j'=1,j' \neq j}^J \rho_{\text{u}} \eta_{\text{u},j'} |h_{\text{u},j'} \vv_j^H \h_\text{u}|^2
    \\
    & \qquad + \sum\nolimits_{k=1}^K \rho_\text{d} \eta_{\text{d},k} |\vv_j^H \h_\text{u} \h_\text{d}^T \w_k|^2 + \sigma_\text{r}^2 |\vv_j^H \h_{\text{u}}|^2.
\end{align}
Since both SINRs have the same structure, we can develop a common method that applies equivalently for both DL and UL; hence, we drop the indices $\text{d}/\text{u}$ and $k/j$ in the sequel.

\vspace{-\baselineskip}

\subsection{Maximizing the SINR in DL and UL}
\label{sec:maxSE}

\vspace{-1mm}

With the SINRs expressed as in \eqref{eq:SINR_du}, we can find a globally optimal amplification gain $r_\text{r}$ which maximizes the SINR for a given phase shift $\phi_\text{r}$ and for given precoding/combining vectors $\{ \w_k \}$ and $\{ \vv_j \}$. To this end, first note that by using the quotient rule, the partial derivative of \eqref{eq:SINR_du} with respect to $r_\text{r}$ can be obtained as
\begin{align}
    \label{eq:dSINR_du}
    \dfrac{\partial}{\partial r_\text{r}} \mathtt{SINR} \! = \! \dfrac{(a_1 b_0 - a_0 b_1) \! + \! 2(a_2 b_0 - a_0 b_2) r_\text{r} \! + \! (a_2 b_1 - a_1 b_2) r_\text{r}^2}{(b_0 + b_1 r_\text{r} + b_2 r_\text{r}^2)^2}.
\end{align}
The numerator of \eqref{eq:dSINR_du} is a second-order polynomial in $r_\text{r}$. Thus, with $c_0 = a_1 b_0 - a_0 b_1,\ c_1 = a_2 b_0 - a_0 b_2,\ c_2 = a_2 b_1 - a_1 b_2$,
the stationary points of \eqref{eq:SINR_du} are
\begin{align}
    \label{eq:stat}
    \nonumber
    \dfrac{\partial}{\partial r_\text{r}} \mathtt{SINR} = 0 & \iff c_0 + 2c_1 r_\text{r} + c_2  r_\text{r}^2 = 0
    \\
    & \iff r_\text{r} = \frac{- c_1 \pm \sqrt{c_1^2 - c_0 c_2}}{c_2}.
\end{align}
It follows that the optimal amplification gain is either one of the two stationary points in \eqref{eq:stat}, or one of the boundary points decided by the power constraint \eqref{eq:P_max}; that is, $0$ or $\sqrt{P_\text{max}/(P_\text{in} + \sigma_\text{r}^2)}$. If either of the two candidates in \eqref{eq:stat} falls outside of the feasible set \eqref{eq:P_max}, it is discarded as a potential solution. To find the optimal amplification gain, we insert the remaining candidates into \eqref{eq:SINR_du} and choose the one that maximizes the SINR. In this way, we find a globally optimal solution $r_\text{r}$ for given $\phi_\text{r}$, $\{ \w_k \}$, and $\{ \vv_j \}$. 

Next, we find a near optimal phase shift by repeating the procedure above for a discrete range of $\phi_\text{r}$. Specifically, we use $S$ equally spaced samples in the range $[0, 2\pi]$; namely $\phi_\text{r} \in \{ 2\pi s / S \}_{s=0}^{S-1}$. After iterating through all phase shifts and finding the optimal amplification gain for each given $\phi_\text{r}$, we let $\phi_\text{r}^*$ and $r_\text{r}^*$ be the pair maximizing the SINR. Then, the optimal repeater gain is $\alpha_\text{r}^* = r_\text{r}^* e^{i \phi_\text{r}^*}$.

The procedure to find $\alpha_\text{r}^*$ is simple, efficient, and has a low computational complexity. However, there is one remaining drawback; the precoding and combining vectors $\{ \w_k \}$ and $\{ \vv_j \}$ are fixed and must be chosen independently of the repeater gain. Otherwise, higher order terms of $r_\text{r}$ are introduced in \eqref{eq:SINR_du}. To overcome this, we iteratively optimize the repeater gain for fixed precoding/combining vectors, and then update those vectors using the newly obtained optimal repeater gain. Note that we do not optimize the precoding and combining vectors per se, we just update them according to some predetermined beamforming scheme (for example, maximum ratio precoding/combining in the directions of the composite DL/UL channels $\{ \mybar\g_{\text{d},k} \}$ and $\{ \mybar\g_{\text{u},j} \}$).

\vspace{-\baselineskip}

\subsection{Proposed Method for Repeater Gain Optimization}

\vspace{-1mm}

The steps of our proposed method for repeater gain optimization are:

\begin{enumerate}[leftmargin=1.5em, label={\arabic*)}]
    \item Initialize the precoding and combining vectors $\{ \w_k^{(0)} \}$ and $\{ \vv_j^{(0)} \}$ for some feasible initial value $\alpha_\text{r}^{(0)}$. Let $n=1$.
    \item Follow the procedure in Section \ref{sec:maxSE} to find an optimal $\alpha_\text{r}^*$ for fixed $\{ \w_k^{(n-1)} \}$ and $\{ \vv_j^{(n-1)} \}$. Let $\alpha_\text{r}^{(n)} = \alpha_\text{r}^*$.
    \item Update the precoding and combining vectors with $\alpha_\text{r}^{(n)}$ to obtain $\{ \w_k^{(n)} \}$ and $\{ \vv_j^{(n)} \}$.
    \item Make $n = n + 1$ and repeat from step 2) until convergence.
\end{enumerate}
Note that in each iteration, the DL power constraint and the output power constraint of the repeater, must be updated with the newly obtained precoding vectors $\{ \w_k^{(n)} \}$.

Our algorithm does not provide any theoretical convergence guarantees. However, in our simulations, we run until the incremental SE improvement is smaller than $10^{-5}$, or until the SE decreases between two iterations. This typically requires $5$--$10$ iterations in DL and $2$--$5$ iterations in UL before converging to a stationary point.

\vspace{-\baselineskip}

\section{NUMERICAL EVALUATIONS}

\vspace{-2mm}

In this section, we evaluate our proposed method for repeater gain optimization. We consider the setup from Fig. \ref{Fig:system} with $M=16$ and $K=J=1$. Further, we assume a scenario where both users and the repeater are restricted to be located on the horizontal line between the two APs. Let $d$ (m) denote the position along this line, where $d=0$ corresponds to the cell border. We let the distance from each of the two APs to the cell border be $100$ m. Thus, the DL AP is at $d=-100$ m and the UL AP at $d=100$ m. Also, we let the DL user be at $d=-50$ m and the UL user at $d=50$ m. The repeater is at position $d_\text{r}$ (m), which we will vary in the range $[-150, 150]$ m. 

All channels are i.i.d. Rayleigh fading with large-scale fading coefficients calculated from the 3GPP Urban Microcell model~\cite[Eqs. (37), (38)]{emil20TWC}, where we assume that the APs, the repeater, and the users have elevations $10$ m, $5$ m, and $1$ m, respectively. The normalized DL and UL transmit powers, $\rho_{\text{d}}$ and $\rho_{\text{u}}$, are chosen such that the median SNR at the receiver side in DL and UL is $15$ dB and $5$ dB, respectively; and we use the maximal DL and UL power control. We let $\sigma_\text{r}^2 = 1$, and the maximal output power $P_\text{max}$ is set to $38$ dBm (before normalization) \cite{Alp2025arXiv}. In the phase shift optimization, we let $S=16$. We initialize $\alpha_\text{r}^{(0)} = 0$, and in each iteration we use maximum ratio precoding/combining vectors in the directions of the composite DL/UL channels; that is, $\w_k = \mybar\g_{\text{d},k}^*$ and $\vv_j = \mybar\g_{\text{u},j}$. 

\vspace{-\baselineskip}

\subsection{Achievable SEs in DL and UL with Optimal Repeater Gains}

\vspace{-1mm}

First, we evaluate the achievable SEs in DL and UL with our proposed method for repeater gain optimization. As a baseline, we consider the case without repeater-assistance, which is equivalent to setting the amplification gain to zero, i.e., $r_\text{r} = 0$. From \eqref{eq:SINR_du}, it follows immediately that this gives $\underline{\mathtt{SINR}} = \mathtt{SINR} |_{r_\text{r} = 0} = a_0 / b_0$, where $a_0$ and $b_0$ are given by \eqref{eq:a0_DL} and \eqref{eq:b0_DL} in DL, respectively \eqref{eq:a0_UL} and \eqref{eq:b0_UL} in UL. The corresponding SE is $\log_2(1 + \underline{\mathtt{SINR}})$.

\begin{figure}[t]
    \centering
    \input{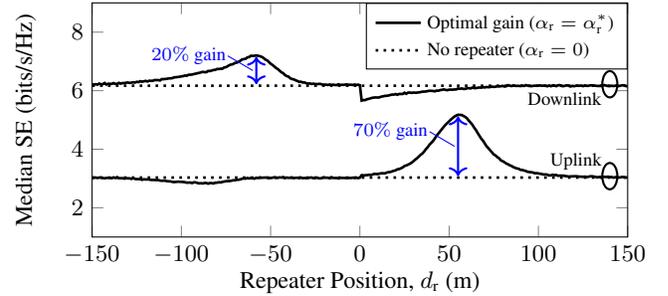}
    \vspace{-8mm}
    \caption{The median DL and UL SEs for a varying repeater position.}
    \label{fig:SE}
    \vspace{-4mm}
\end{figure}

In Fig. \ref{fig:SE}, we compare the median DL and UL SEs achieved with our proposed repeater gain optimization (solid lines), to the baseline without repeater-assistance (dotted lines). We vary the repeater position in the range $d_\text{r} \in [-150, 150]$ m, while the AP and user positions are fixed (see above). The repeater assists the cell in which it is located, i.e., for $d_\text{r} < 0$ the repeater gain is optimized to maximize the SE for the DL user and for $d_\text{r} > 0$ it is optimized for the UL user.

We note that the repeater can boost performance by up to, roughly, $20 \%$ in DL and $70 \%$ in UL, and that these maximums are obtained at $d_\text{r} = -58$ m and $d_\text{r} = 55$ m, respectively. This indicates that the repeater should be located in the vicinity of the users, but preferably closer to the AP than the cell border, to maximize the gains. This is an intuitive result; when moving the repeater closer to the cell border, it picks up more CLI and less of the desired signal, leading to diminishing SE gains. We also note that the user for which the repeater is not optimized can experience a slight SE reduction; however, this loss is much less prominent than the gains.

\vspace{-\baselineskip}

\subsection{Real-Valued Versus Complex-Valued Repeater Gains}

\vspace{-1mm}

Next, we evaluate the benefit of using a complex-valued repeater gain, i.e., of optimizing the repeater phase shift in addition to the amplification. As a baseline, we restrict the repeater gain to be real-valued, which corresponds to the special case of our proposed method with $S=1$, i.e., we only consider $\phi_\text{r} = 0$. We also include the performance of the baseline without repeater-assistance.

\begin{figure}[t]
    \centering
    \input{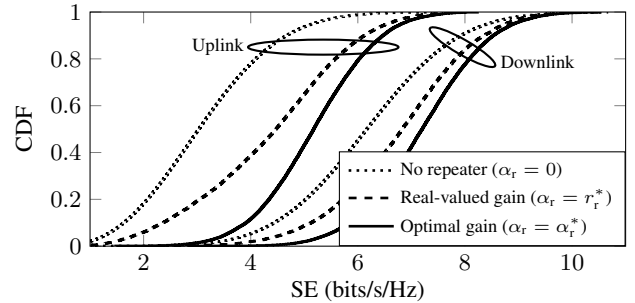}
    \vspace{-\baselineskip}
    \caption{DL and UL SEs with optimal and real-valued repeater gains.}
    \label{fig:gain}
    \vspace{-6mm}
\end{figure}

In Fig. \ref{fig:gain}, we show the CDFs of the achievable DL and UL SEs at the repeater positions for which we obtained the maximal improvement in Fig. \ref{fig:SE}. That is, for $d_\text{r} = -58$ m in DL and $d_\text{r} = 55$ m in UL. We note that only optimizing the (real-valued) amplification gain gives a noticeable improvement over the case with no repeater. However, the gain is even more significant when jointly optimizing the phase shift, especially for users experiencing relatively low SEs, whereas the gain is smaller at the peak SEs. 

\vspace{-\baselineskip}

\section{CONCLUSIONS AND FUTURE DIRECTIONS}

\vspace{-2mm}

This paper has investigated {how much dynamic TDD MIMO systems can benefit from repeater-assistance, considering that repeaters amplify the CLI in addition to the desired signal.} In most cases, our proposed repeater gain optimization framework based on \gls{dl} and \gls{ul} SE maximization provides improved \gls{dl} and \gls{ul} SEs compared to the case when the repeater is deactivated. Specifically, we have identified conditions on the repeater location  under which such improvements in the \gls{dl} and \gls{ul} SEs can be obtained. Future work may investigate setups with multiple repeaters, and a joint optimization of the repeater gains along with \gls{dl}/\gls{ul} power control and precoding/combining vectors, which might further improve performance and overcome the cases where the repeater reduces performance.

\clearpage

\bibliographystyle{IEEEtran}
\bibliography{refs}

\end{document}